\documentclass[conference]{IEEEtran}
\usepackage[pdftex]{graphicx}
\usepackage[usenames,dvipsnames,table,xcdraw]{xcolor}
\usepackage{array}
\usepackage{multirow}
\usepackage[normalem]{ulem}
\useunder{\uline}{\ul}{}
\usepackage{booktabs}
\usepackage{supertabular,booktabs}
\usepackage{longtable}
\usepackage{lscape}
\usepackage{cite}
\usepackage{url}
\usepackage{hyperref}
\usepackage{dirtytalk}
\usepackage{graphicx}
\usepackage{comment}
\usepackage{makecell}
\usepackage{svg}
\usepackage{amsmath}

\usepackage[caption=false]{subfig}
\graphicspath{{Images/}}
\usepackage[normalem]{ulem}
\useunder{\uline}{\ul}{}

\usepackage{tabularx}
\usepackage{balance}
\usepackage{nameref}

\begin{document}

\title{Towards User-Focused Cross-Domain Testing: Disentangling Accessibility, Usability, and Fairness}

\author{
\IEEEauthorblockN{Matheus de Morais Leça}
\IEEEauthorblockA{University of Calgary\\
Calgary, AB, Canada \\
matheus.demoraisleca@ucalgary.ca}
\and

\IEEEauthorblockN{Ronnie de Souza Santos}
\IEEEauthorblockA{University of Calgary\\
Calgary, AB, Canada \\
ronnie.desouzasantos@ucalgary.ca}

}


\IEEEtitleabstractindextext{%
\begin{abstract}
Fairness testing is increasingly recognized as fundamental in software engineering, especially in the domain of data-driven systems powered by artificial intelligence. However, its practical integration into software development may pose challenges, given its overlapping boundaries with usability and accessibility testing. In this tertiary study, we explore these complexities using insights from 12 systematic reviews published in the past decade, shedding light on the nuanced interactions among fairness, usability, and accessibility testing and how they intersect within contemporary software development practices.
\end{abstract}

}

\maketitle

\IEEEdisplaynontitleabstractindextext

\IEEEpeerreviewmaketitle

\section{Introduction}
\label{sec:introduction}

As technology continues to play an increasingly integral role in our lives, the expectation that software is effective in representing the multifaceted characteristics of our society becomes not only a technical necessity but also an ethical obligation for developing software that is both equitable and inclusive \cite{albusays2021diversity}. In this context, software fairness emerges as a non-functional requirement and a quality attribute of systems, especially those reliant on data-driven processes and powered by artificial intelligence, to ensure just and unbiased treatment of users or groups of users affected by the outcomes of algorithms and applications \cite{brun2018software, zhang2021ignorance, galhotra2017fairness}.

While not entirely novel in software development, the concept of software fairness has recently garnered substantial attention. Historically, software development focused on functionality and performance, often overlooking potential biases and ethical implications in algorithms. Yet, the surge in discussions on software engineering for artificial intelligence and ethics in machine learning has brought software fairness into the spotlight, reflecting a shift toward recognizing its essential role in modern software engineering practices \cite{vera2017may, brun2018software, galhotra2017fairness}.

As a quality attribute, software fairness intersects with other essential software characteristics, namely usability and accessibility. Similar to usability and accessibility, fairness profoundly impacts user experiences. However, a potential challenge stemming from these parallels is the assumption that fairness-related requirements, design decisions, or validations have been adequately addressed alongside these other software elements traditionally associated with accessibility or usability \cite{ryan2023integrating, shin2019role}. This scenario led to the following research question: \textit{\textbf{What is the interplay among accessibility, usability, and fairness testing in software development?}} 

Our motivation to explore this matter is rooted in establishing fairness testing as a foundational practice in software development. It is crucial to distinguish its specific characteristics from other longstanding testing methods. This distinction is essential for comprehending the nuanced dynamics of systemic bias and for promoting a more inclusive, ethical, and effective approach to software system development.

The remainder of this paper is organized as follows.  In Section~\ref{sec:background}, we present definitions and previous work that guided this research. In Section~\ref{sec:method}, we present our method. In Section~\ref{sec:findings}, we introduce the key insights identified so far. In Section~\ref{sec:discuss}, we discuss the implications of our findings and opportunities for future research. Finally, Section~\ref{sec:conclusions} summarizes our contributions and final considerations.

\section{Background} \label{sec:background}
Software testing is a dynamic and essential process that ensures software products align with client and user expectations by verifying the correct implementation of all planned features, behaviors, and quality requirements. Testing is necessary to ensure reliability, functionality, and performance under diverse conditions and to identify defects. As software complexity and user demands evolve, robust testing practices remain indispensable for delivering products that are resilient, user-friendly, and aligned with industry standards ~\cite{abran2001guide, bertolino2007software}.

Software testing encompasses a diverse array of methodologies aimed at ensuring the quality and functionality of software products. Many of these methodologies are geared toward verifying how a system interacts with users to ensure a positive experience \cite{abran2001guide}. Key examples include usability testing, which evaluates the software's ease of use and user interface design \cite{mason2005critical, bandi2013usability}; accessibility testing, which ensures the software can be used effectively by people with disabilities \cite{stray2019empowering}; and emergent fairness testing, which verifies whether the software behaves ethically and without bias, particularly in AI-driven systems \cite{galhotra2017fairness}.

Usability testing evaluates the ease of use and user-friendliness of a software application or system. Its primary goal is to identify usability issues like navigation challenges, confusing interface elements, or inefficient workflows that could impact user experience negatively. In this process, software testers can conduct evaluations and heuristic reviews based on established usability principles, as well as set up controlled testing environments to observe user behavior, gather feedback, and iterate on the software to enhance usability and overall user satisfaction \cite{mason2005critical, bandi2013usability, zhang2005challenges}.

Accessibility testing evaluates the extent to which a software application or system can be used effectively by people with disabilities. It focuses on ensuring that individuals with diverse abilities, including those with visual, auditory, motor, or cognitive impairments, can access and interact with the software without major difficulties. In this process, testers employ various techniques and tools to identify and address issues that hinder users with disabilities from having a positive experience using the software, i.e., to make the software inclusive and accessible to all users, regardless of their abilities \cite{stray2019empowering, bai2018categorization, sanchez2017method}.

Fairness testing examines the ethical implications and biases within software systems, particularly those powered by artificial intelligence. Testers use specialized techniques and tools to assess whether the software's decisions or outputs exhibit fairness across different user groups. This process involves identifying and mitigating biases that could result in discriminatory outcomes based on factors such as race, gender, or socioeconomic status. Fairness testing aims to ensure that the software treats all users equally and without bias \cite{galhotra2017fairness, aggarwal2019black}.
\section{Method} \label{sec:method}
In this research, we explored published evidence on usability, accessibility, and fairness testing. This evidence was collected from previous systematic literature reviews, categorizing this study as a tertiary study \cite{kitchenham2010systematic, zein2023systematic}. A tertiary study is a type of systematic review that addresses broader research questions by identifying and analyzing previous systematic reviews, using a methodology similar to systematic mapping studies \cite{zein2023systematic}. Unlike traditional systematic literature reviews, which consolidate outcomes from primary experimental studies (such as experiments, case studies, or surveys) \cite{kitchenham2004evidence}, a tertiary study focuses on synthesizing findings from secondary studies—specifically, systematic reviews and mapping studies \cite{da2010six}.

\subsection{Search Strategy}
Following the guidelines for conducting systematic reviews in software engineering \cite{kitchenham2004evidence} and drawing insights from previous tertiary studies \cite{kitchenham2010systematic, zein2023systematic}, we conducted a search for secondary studies specifically focusing on usability testing, accessibility testing, and fairness testing. In this process, we employed a semi-automated approach to identify papers for our review, deviating from the typical structured search string used in systematic reviews. Instead, we conducted a straightforward search using the term \textit{systematic review} in combination with either \textit{usability testing}, \textit{accessibility testing}, or \textit{fairness testing}. This strategy aimed to rapidly filter out studies that mention these terms but fall outside the scope of our goal. Additionally, to ensure relevance, the studies retrieved in our search were immediately evaluated to determine whether they fell within the scope of our investigation. Our focus was specifically on systematic reviews addressing usability testing, accessibility testing, or fairness testing. We employed a two-step approach to filter the relevant papers to streamline this process. First, we conducted a rapid screening by scanning each paper's title, abstract, and conclusion to identify those most likely to align with our research focus. Next, we thoroughly reviewed the full text of the remaining papers to confirm their direct relevance to the topics under investigation. The search was conducted across leading scientific databases, including IEEE Xplore, ACM Digital Library, SpringerLink, and Google Scholar. These databases were selected for their reputation, comprehensive coverage, and relevance in providing high-quality research papers on the topic under investigation.

Subsequently, we conducted a manual search for studies presented at several of the world's leading conferences on software engineering and software testing. These conferences included the International Conference on Software Engineering, the International Conference on Software Testing, the International Conference on Software Maintenance and Evolution, and the International Conference on the Foundations of Software Engineering. Our selection was guided by their reputation as premier venues in the field, rigorous peer-review processes, and focus on advancing state-of-the-art research. These conferences also serve as essential platforms for discussions on topics directly aligned with our research focus, ensuring relevance and credibility in the studies included. Our search focused on publications from the past decade, specifically between 2015 and 2024, recognizing the increasing prominence of discussions on software fairness within software engineering. 

\subsection{Study Selection}
For each topic, the search process was conducted separately. We analyzed the retrieved results based on the paper titles and abstracts. In this tertiary study, we included papers that reported systematic reviews or mapping studies focused on one of the three types of testing or had a broader focus that included discussions on testing. We excluded systematic reviews or mapping studies that did not have usability testing, accessibility testing, or fairness testing as the central focus of their findings or discussions.

\subsection{Data analysis}
Excerpts were extracted directly from the studies and recorded in a spreadsheet to facilitate data analysis. Primarily, we collected qualitative data, including definitions, characteristics, and aspects of each type of testing. Consequently, qualitative analysis was employed to thoroughly examine the data and provide material for our discussion. In this process, we applied thematic analysis \cite{cruzes2011recommended} to comprehensively explore and interpret the collected data.

\section{Findings} 
\label{sec:findings}
After completing the search and applying the filtration process, we identified 12 systematic literature reviews published in the past decade. Six of them were conducted in the context of usability testing, three focused on accessibility testing, and three discussed fairness testing. Table \ref{table:distribution} presents a summary of our identified and analyzed reviews.
\begin{table}[!ht]

\footnotesize
\caption{Systematic Reviews Analysed in the Tertiary Study}
\label{table:distribution}
\begin{tabularx}{\linewidth}{p{1.2cm} p{5.5cm} X}
\midrule
\textbf{Aspect} & \textbf{Study} & \textbf{Year}\\
\midrule

\textbf{Usability}
& Virtual Reality on Product Usability Testing: A Systematic Literature Review \cite{freitas2020virtual} & 2020 \\
& Usability Evaluation Methods of Mobile Applications: A Systematic Literature Review \cite{nugroho2022usability} & 2022 \\
& Reporting Usability Defects: A Systematic Literature Review \cite{yusop2016reporting} & 2016 \\
& Usability of mobile learning applications: a systematic literature review \cite{kumar2018usability} & 2018 \\
& Empirical Studies on Usability of mHealth Apps: A Systematic Literature Review \cite{zapata2015empirical}  & 2015\\
& How to develop usability heuristics: A systematic literature review \cite{quinones2017develop} & 2017 \\
\midrule 

\textbf{Accessibility} & Accessibility engineering in web evaluation process: a systematic literature review \cite{ara2023accessibility} & 2023 \\
& Accessible Features to Support Visually Impaired People in Game Development: A Systematic Literature Review \cite{garcez2020accessible}  & 2020\\
& Web accessibility evaluation methods: a systematic review \cite{nunez2019web} & 2019 \\
\midrule 

\textbf{Fairness} & Software Fairness: An Analysis and Survey \cite{soremekun2022software} & 2022 \\
& Fairness Testing: A Comprehensive Survey and Analysis of
Trends \cite{chen2024fairness} & 2023 \\
& Fairness perceptions of algorithmic decision-making: A systematic review of the empirical literature \cite{starke2022fairness} & 2022 \\
\bottomrule
\end{tabularx}
\end{table}
\vspace{-2.5mm}

\subsection{Usability, Accessibility and Fairness Definitions}
We explored the definitions presented in the systematic reviews and the characteristics attributed to each type of testing to examine the interplay among the concepts of usability, accessibility, and fairness. Table \ref{tab:concepts} presents the various definitions of each concept identified in the papers. As expected, the literature reviews on software fairness are more recent than those on usability and accessibility. Additionally, we found more reviews on usability than on accessibility. One reason for this discrepancy is that, in many instances, these two aspects of software development were investigated together, with the focus shifting between them. This reinforces our motivation to discuss their interplay. Analyzing this evidence, we can emphasize the following: \\

\noindent\textbf{Usability}. The concept of usability testing is closely tied to how easily users can interact with the system and how the software enriches their experience. Key elements associated with usability in the reviews included interface, pleasantness, learnability, understandability, and ease of use. Usability defects were often associated with poor navigation, inconsistent interface elements, and problems with layout. 
Although the authors of this study acknowledge the distinctions between usability and user experience (UX) within the context of software engineering, we will adhere to the definitions provided in these literature reviews. For consistency, we will use the term \textit{usability} to encompass user experience and any form of user interaction with a solution. \\

\noindent\textbf{Accessibility}. The concept of accessibility testing is closely related to a specific context focused on providing support or a particular software configuration. Key elements associated with accessibility in the reviews included support for disabilities, accommodations for impairment, and universality. Accessibility defects were often associated with a lack of adaptable elements to support users, such as text, audio, colors, and images. \\

\noindent\textbf{Fairness}. The concept of fairness testing is closely related to how systems treat users and the need for them to be free from biased outcomes. Key elements associated with fairness in the reviews encompassed a decrease in discrimination, treatment of minorities, and ethical considerations. Fairness defects were often associated with data handling, leading to biased outcomes, such as discrimination, limitation, and exclusion. This breakdown highlights the distinct yet interconnected nature of usability, accessibility, and fairness in software development, reflecting their critical roles in enhancing user experience, inclusivity, and ethical standards within software products. \\

\subsection{Usability, Accessibility, and Fairness Testing Interplay}

The data collected from the reviews demonstrate an intricate interplay among usability, accessibility, and fairness in software development. Traditionally, usability and accessibility have been closely linked, particularly through interface elements that facilitate effective and user-friendly interactions for different individuals. Usability testing focuses on user experience and satisfaction, while accessibility testing ensures essential support through interface design tailored to diverse needs. Now, fairness introduces a new dimension by testing for equitable outcomes derived from data and algorithmic decisions.

In our analysis, we observed a strong correlation between fairness testing and accessibility testing, especially because individuals with disabilities often belong to minority groups. Accessibility testing rigorously evaluates features to accommodate diverse impairments, such as visual or speech impairments. In contrast, fairness testing addresses complexities involving intersecting biases affecting various underrepresented groups, including people with disabilities. Furthermore, there is a significant relationship between usability testing and fairness testing. While usability testing aims to enhance user experience during software interaction, fairness testing ensures equitable interaction across different user groups, thereby preventing exclusions or limited experiences based on user background (e.g., gender, sexual orientation, ethnicity).

Based on the evidence analyzed from the literature, it is evident that usability, accessibility, and fairness play intricate and multifaceted roles in software development. Usability focuses on creating effective and enjoyable interfaces for a wide range of users, while accessibility ensures these interfaces are usable by individuals with disabilities. Fairness, on the other hand, strives to achieve unbiased outcomes, interfaces, and interactions across diverse user groups. This interplay demonstrates a potential challenge in integrating fairness testing as a routine practice in the software development process, highlighting the need for the development of strategies, methods, and tools to systematically incorporate fairness testing into the software industry.

Additionally, the relationship between usability, accessibility, and fairness creates complex challenges in identifying and fixing fairness defects. The combined effect of these usability and accessibility problems can divert attention away from fairness, delaying necessary interventions and perpetuating biases within the software. If testers are solely focused on usability problems, they may not uncover the parts of the software where fairness issues exist, masking the urgency to address these defects. Similarly, accessibility bugs, such as inadequate support for assistive technologies or non-compliance with accessibility standards, can hinder software testing from addressing other types of inclusivity issues, making it difficult to identify and prioritize these concerns.

\begin{table}[!ht]
\footnotesize
\caption{Usability, Accessibility, and Fairness Characteristics}
\label{tab:concepts}
\begin{tabularx}{\linewidth}{p{1.2cm} p{1.7cm} X}
\toprule
\textbf{Category} & \textbf{Subcategory} & \textbf{Illustrative Quotation} \\
\midrule

\textbf{Usability} 
& User Friendly
& ``(...) the extent to which a system, product, or service can be used by specified users to achieve specified goals with effectiveness, efficiency, and satisfaction.'' \\
& Pleasant 
& ``(...) an unintended behavior by the product that is noticed by the user and has an effect on the user experience.'' \\
& Learnable 
& ``(...) a multiple properties of a user interface (...): Learnability (easy to be learned), Memorability (easy to be remembered).'' \\
& Understandable 
& ``Usable software should not only have an attractive user interface, but it should be easy to understand, learn, operate and also control.'' \\
& Ease-to-Use 
& ``Perceived ease of use refers to the degree to which the user expects that using a particular service would be free of effort.'' \\
\midrule

\textbf{Accessibility} 
& Disability-Inclusive 
& ``(...) the degree to which data can be accessed (...), particularly by people who need supporting technology or specific configuration because of some disability.'' \\
& Impairment-Inclusive 
& ``(...) requirement that establishes how software and computers can be made accessible to users with various types of impairments.'' \\
& Universal 
& ``(...) people with special needs should be able to access, navigate, interact, and contribute to the information that is available.'' \\
\midrule

\textbf{Fairness} 
& Unbiased 
& ``(...) a property of learning-based systems which aims to ensure that the software does not exhibit biases.'' \\
& Minority-Inclusive 
& ``(...) requires software to treat different demographic groups in a similar manner.'' \\
& Ethically Aware 
& ``(...) embedding societal values in the design of ADM systems.'' \\
\bottomrule
\end{tabularx}
\end{table}
\vspace{-2.5mm}

\begin{figure*}
    \centering
    \includegraphics[width=\linewidth]{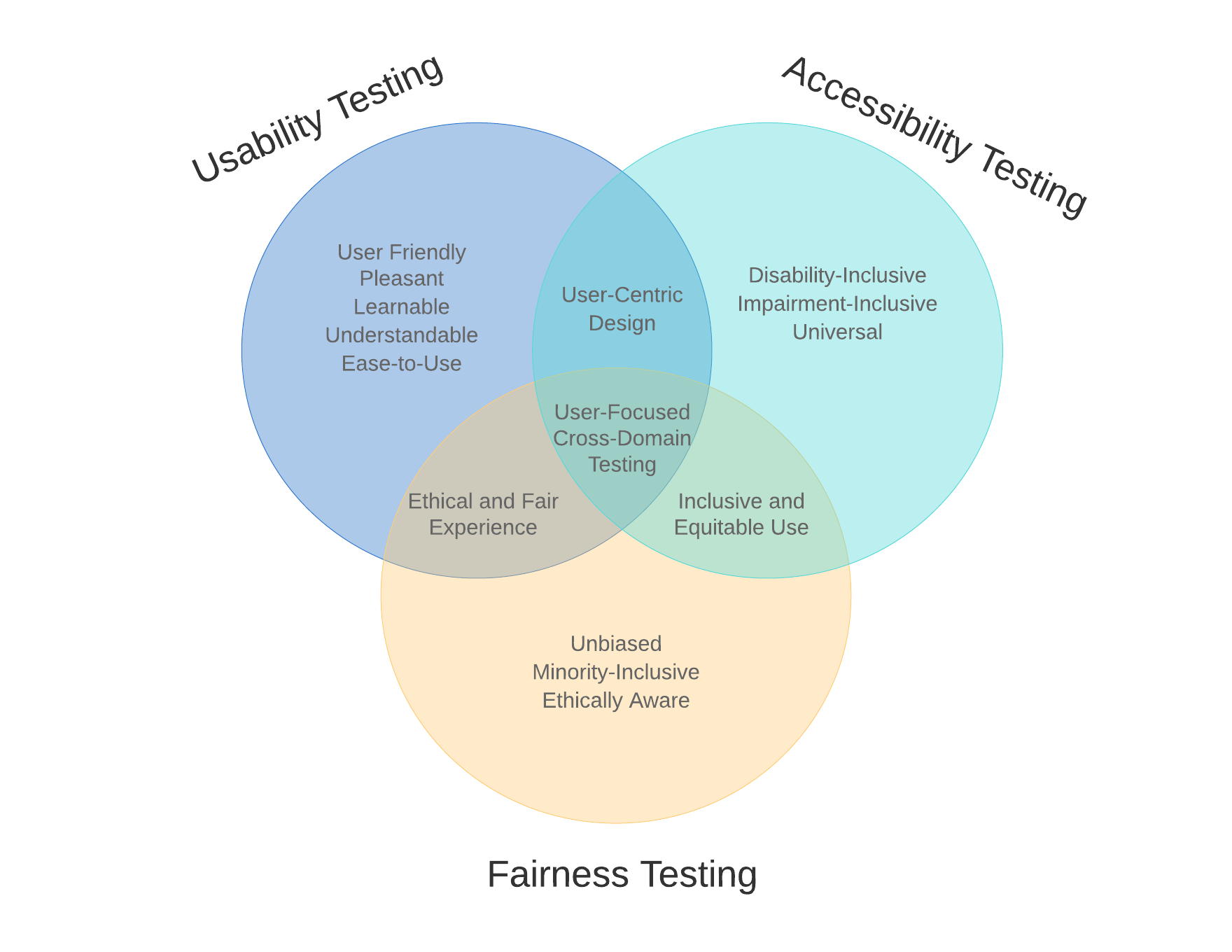}
    \caption{Holistic Testing Perspective on Usability, Accessibility, and Fairness}
    \label{fig:holistic}
\end{figure*}

\section{Discussions} \label{sec:discuss}

Our study synthesizes literature to explore discussions surrounding fairness testing, investigating how distinctions between usability, accessibility, and fairness can become blurred, thereby posing challenges in the testing process. As a key point, we observed that while usability and accessibility adhere to established ISO norms and global guidelines frequently cited in studies, fairness in software development is still evolving. Through our analysis, we caution against the misconception that pursuing fairness involves merely refining overall system usability by integrating ethical principles and addressing data biases into usability and accessibility testing. Right now, it is essential to avoid this oversimplification when establishing fairness testing as an industry practice.

This involves recognizing software fairness as integral throughout the entire software development process rather than treating it as a secondary concern. Adopting this perspective can help us better meet the needs of a diverse user base and start addressing the challenges associated with implementing solutions that proactively mitigate biases, taking into account various underrepresented groups in our society. Our study takes a preliminary step towards this goal by exploring the intersections and complexities among fairness, usability, and accessibility in the context of software testing.

\subsection{A Holistic Software Testing Perspective on Usability, Accessibility, and Fairness}

Usability testing and accessibility testing overlap in their commitment to user-centric design. This intersection ensures that interfaces are not only easy to use but also inclusive, accommodating accessibility needs such as support for disabilities and universality. By integrating these aspects, the software can be tested to guarantee that it serves a diverse user base, ensuring usability features are accessible to everyone, including those with impairments.

Another overlap exists between usability testing and fairness testing, emphasizing the importance of verifying and validating an equitable user experience that incorporates ethical considerations into design decisions. Usability focuses on delivering a pleasant and effective user experience, ensuring interfaces are intuitive and enjoyable, while fairness adds another layer by ensuring these interactions are ethical, providing a positive experience for all users regardless of their background or identity.

The overlap between accessibility and fairness shifts the focus of testing to equitable access and inclusion. Accessibility features are instrumental in promoting fairness and equity by ensuring software is universally accessible. This overlap ensures that bias within the software does not lead to discrimination against users based on disabilities or minority status, promoting equal access and opportunities.

At the heart of these relationships lies a central overlap that integrates usability, accessibility, and fairness into a user-focused, cross-domain testing approach (Figure \ref{fig:holistic}). This approach combines principles from all three domains to ensure the development of software that is user-friendly, inclusive, fair, and ethically sound. By adopting this comprehensive approach, the software industry can produce systems that cater to the varied needs of users effectively while upholding principles of fairness and inclusivity in digital environments.



\subsection{Limitations} \label{sec:limitations}
We acknowledge limitations in this study inherent to tertiary research. The primary challenge in this type of research lies in correctly identifying all relevant systematic reviews for analysis. In this regard, we understand that a significant number of systematic reviews might not have been included in our study because our semi-automatic search method was designed with the sole objective of pinpointing recent reviews to assist in the process of reflecting and discussing the intricate relationships among usability testing, accessibility testing, and fairness testing. Despite this limitation, this study represents an essential initial step in shedding light on the blurred lines within these aspects of software testing.


\section{Conclusions} 
\label{sec:conclusions}
In this study, our primary goal was to explore the interplay among usability testing, accessibility testing, and fairness testing, emphasizing the need for a nuanced understanding of their distinct characteristics. To support our investigation, we conducted an analysis of systematic literature reviews published in the past decade. These reviews provided us with definitions and general characteristics of software usability, accessibility, and fairness, enabling us to compare and contrast their roles in software testing.

Our findings demonstrate that the boundaries between these three types of tests are often blurred, which can lead to misconceptions and misinterpretations. In particular, fairness testing risks are being overlooked throughout the software development lifecycle, as currently, there are no established methods or guidelines to guide software professionals in implementing it. To address these challenges, elevating fairness to a first-class entity deserving distinct attention and strategic testing is essential. 

Recognizing and exploring this interplay has the potential to promote a more inclusive, equitable, and effective approach to software testing. Hence, this study marks an initial step toward enhancing our understanding of these critical aspects. Moving forward, future research should focus on refining methodologies for integrating fairness testing into standard software development practices, thereby fostering environments that prioritize fairness alongside usability and accessibility.

\section*{Acknowledgments}
This project was supported by the NSERC Discovery Grant RGPIN-2024-06260.

\ifCLASSOPTIONcaptionsoff
  \newpage
\fi

\balance
\bibliographystyle{IEEEtran}
\bibliography{bib.bib}

\begin{thebibliography}{10}
\providecommand{\url}[1]{#1}
\csname url@samestyle\endcsname
\providecommand{\newblock}{\relax}
\providecommand{\bibinfo}[2]{#2}
\providecommand{\BIBentrySTDinterwordspacing}{\spaceskip=0pt\relax}
\providecommand{\BIBentryALTinterwordstretchfactor}{4}
\providecommand{\BIBentryALTinterwordspacing}{\spaceskip=\fontdimen2\font plus
\BIBentryALTinterwordstretchfactor\fontdimen3\font minus \fontdimen4\font\relax}
\providecommand{\BIBforeignlanguage}[2]{{%
\expandafter\ifx\csname l@#1\endcsname\relax
\typeout{** WARNING: IEEEtran.bst: No hyphenation pattern has been}%
\typeout{** loaded for the language `#1'. Using the pattern for}%
\typeout{** the default language instead.}%
\else
\language=\csname l@#1\endcsname
\fi
#2}}
\providecommand{\BIBdecl}{\relax}
\BIBdecl

\bibitem{albusays2021diversity}
K.~Albusays, P.~Bjorn, L.~Dabbish, D.~Ford, E.~Murphy-Hill, A.~Serebrenik, and M.-A. Storey, ``The diversity crisis in software development,'' \emph{IEEE Software}, vol.~38, no.~2, pp. 19--25, 2021.

\bibitem{brun2018software}
Y.~Brun and A.~Meliou, ``Software fairness,'' in \emph{Proceedings of the 2018 26th ACM joint meeting on european software engineering conference and symposium on the foundations of software engineering}, 2018, pp. 754--759.

\bibitem{zhang2021ignorance}
J.~M. Zhang and M.~Harman, ``" ignorance and prejudice" in software fairness,'' in \emph{2021 IEEE/ACM 43rd International Conference on Software Engineering (ICSE)}.\hskip 1em plus 0.5em minus 0.4em\relax IEEE, 2021, pp. 1436--1447.

\bibitem{galhotra2017fairness}
S.~Galhotra, Y.~Brun, and A.~Meliou, ``Fairness testing: testing software for discrimination,'' in \emph{Proceedings of the 2017 11th Joint meeting on foundations of software engineering}, 2017, pp. 498--510.

\bibitem{vera2017may}
M.~Vera, A.~M. Rodr{\'\i}guez-S{\'a}nchez, and M.~Salanova, ``May the force be with you: Looking for resources that build team resilience,'' \emph{Journal of Workplace Behavioral Health}, vol.~32, no.~2, pp. 119--138, 2017.

\bibitem{ryan2023integrating}
S.~Ryan, C.~Nadal, and G.~Doherty, ``Integrating fairness in the software design process: An interview study with hci and ml experts,'' \emph{IEEE Access}, vol.~11, pp. 29\,296--29\,313, 2023.

\bibitem{shin2019role}
D.~Shin and Y.~J. Park, ``Role of fairness, accountability, and transparency in algorithmic affordance,'' \emph{Computers in Human Behavior}, vol.~98, pp. 277--284, 2019.

\bibitem{abran2001guide}
A.~Abran, P.~Bourque, R.~Dupuis, and J.~W. Moore, \emph{Guide to the software engineering body of knowledge-SWEBOK}.\hskip 1em plus 0.5em minus 0.4em\relax IEEE Press, 2001.

\bibitem{bertolino2007software}
A.~Bertolino, ``Software testing research: Achievements, challenges, dreams,'' in \emph{Future of Software Engineering (FOSE'07)}.\hskip 1em plus 0.5em minus 0.4em\relax IEEE, 2007, pp. 85--103.

\bibitem{mason2005critical}
P.~Mason and B.~Plimmer, ``A critical comparison of usability testing methodologies,'' \emph{NACCQ (Tauranga)}, pp. 255--258, 2005.

\bibitem{bandi2013usability}
A.~Bandi and P.~Heeler, ``Usability testing: A software engineering perspective,'' in \emph{2013 International Conference on Human Computer Interactions (ICHCI)}.\hskip 1em plus 0.5em minus 0.4em\relax IEEE, 2013, pp. 1--8.

\bibitem{stray2019empowering}
V.~Stray, A.~Bai, N.~Sverdrup, and H.~Mork, ``Empowering agile project members with accessibility testing tools: a case study,'' in \emph{Agile Processes in Software Engineering and Extreme Programming: 20th International Conference, XP 2019, Montr{\'e}al, QC, Canada, May 21--25, 2019, Proceedings 20}.\hskip 1em plus 0.5em minus 0.4em\relax Springer International Publishing, 2019, pp. 86--101.

\bibitem{zhang2005challenges}
D.~Zhang and B.~Adipat, ``Challenges, methodologies, and issues in the usability testing of mobile applications,'' \emph{International journal of human-computer interaction}, vol.~18, no.~3, pp. 293--308, 2005.

\bibitem{bai2018categorization}
A.~Bai, K.~Fuglerud, R.~A. Skjerve, and T.~Halbach, ``Categorization and comparison of accessibility testing methods for software development,'' \emph{Transforming our World Through Design, Diversity and Education}, pp. 821--831, 2018.

\bibitem{sanchez2017method}
S.~Sanchez-Gordon and S.~Luj{\'a}n-Mora, ``A method for accessibility testing of web applications in agile environments,'' in \emph{Proceedings of the 7th World Congress for Software Quality (WCSQ). En proceso de publicaci{\'o}n.(citado en la p{\'a}gina 13, 15, 85)}, 2017, p. 144.

\bibitem{aggarwal2019black}
A.~Aggarwal, P.~Lohia, S.~Nagar, K.~Dey, and D.~Saha, ``Black box fairness testing of machine learning models,'' in \emph{Proceedings of the 2019 27th ACM joint meeting on european software engineering conference and symposium on the foundations of software engineering}, 2019, pp. 625--635.

\bibitem{kitchenham2010systematic}
B.~Kitchenham, R.~Pretorius, D.~Budgen, O.~P. Brereton, M.~Turner, M.~Niazi, and S.~Linkman, ``Systematic literature reviews in software engineering--a tertiary study,'' \emph{Information and software technology}, vol.~52, no.~8, pp. 792--805, 2010.

\bibitem{zein2023systematic}
S.~Zein, N.~Salleh, and J.~Grundy, ``Systematic reviews in mobile app software engineering: A tertiary study,'' \emph{Information and Software Technology}, vol. 164, p. 107323, 2023.

\bibitem{kitchenham2004evidence}
B.~A. Kitchenham, T.~Dyba, and M.~Jorgensen, ``Evidence-based software engineering,'' in \emph{Proceedings. 26th International Conference on Software Engineering}.\hskip 1em plus 0.5em minus 0.4em\relax IEEE, 2004, pp. 273--281.

\bibitem{da2010six}
F.~da~Silva, A.~Santos, S.~Soares, A.~Fran{\c{c}}a, and C.~Monteiro, ``Six years of systematic literature reviews in software engineering: an extended tertiary study,'' in \emph{32th international conference on software, ICSE}, vol.~10, 2010, pp. 1--10.

\bibitem{cruzes2011recommended}
D.~S. Cruzes and T.~Dyba, ``Recommended steps for thematic synthesis in software engineering,'' in \emph{2011 international symposium on empirical software engineering and measurement}.\hskip 1em plus 0.5em minus 0.4em\relax IEEE, 2011, pp. 275--284.

\bibitem{freitas2020virtual}
F.~Freitas, H.~Oliveira, I.~Winkler, and M.~Gomes, ``Virtual reality on product usability testing: A systematic literature review,'' in \emph{2020 22nd Symposium on Virtual and Augmented Reality (SVR)}.\hskip 1em plus 0.5em minus 0.4em\relax IEEE, 2020, pp. 67--73.

\bibitem{nugroho2022usability}
A.~Nugroho, P.~I. Santosa, and R.~Hartanto, ``Usability evaluation methods of mobile applications: A systematic literature review,'' in \emph{2022 International Symposium on Information Technology and Digital Innovation (ISITDI)}.\hskip 1em plus 0.5em minus 0.4em\relax IEEE, 2022, pp. 92--95.

\bibitem{yusop2016reporting}
N.~S.~M. Yusop, J.~Grundy, and R.~Vasa, ``Reporting usability defects: A systematic literature review,'' \emph{IEEE Transactions on Software Engineering}, vol.~43, no.~9, pp. 848--867, 2016.

\bibitem{kumar2018usability}
B.~A. Kumar and P.~Mohite, ``Usability of mobile learning applications: a systematic literature review,'' \emph{Journal of Computers in Education}, vol.~5, pp. 1--17, 2018.

\bibitem{zapata2015empirical}
B.~C. Zapata, J.~L. Fern{\'a}ndez-Alem{\'a}n, A.~Idri, and A.~Toval, ``Empirical studies on usability of mhealth apps: a systematic literature review,'' \emph{Journal of medical systems}, vol.~39, pp. 1--19, 2015.

\bibitem{quinones2017develop}
D.~Qui{\~n}ones and C.~Rusu, ``How to develop usability heuristics: A systematic literature review,'' \emph{Computer standards \& interfaces}, vol.~53, pp. 89--122, 2017.

\bibitem{ara2023accessibility}
J.~Ara, C.~Sik-Lanyi, and A.~Kelemen, ``Accessibility engineering in web evaluation process: a systematic literature review,'' \emph{Universal Access in the Information Society}, pp. 1--34, 2023.

\bibitem{garcez2020accessible}
L.~Garcez, M.~Thiry, and A.~Fernandes, ``Accessible features to support visually impaired people in game development:: A systematic literature review of the last 15 years,'' in \emph{2020 15th Iberian Conference on Information Systems and Technologies (CISTI)}.\hskip 1em plus 0.5em minus 0.4em\relax IEEE, 2020, pp. 1--6.

\bibitem{nunez2019web}
A.~Nu{\~n}ez, A.~Moquillaza, and F.~Paz, ``Web accessibility evaluation methods: a systematic review,'' in \emph{Design, User Experience, and Usability. Practice and Case Studies: 8th International Conference, DUXU 2019, Held as Part of the 21st HCI International Conference, HCII 2019, Orlando, FL, USA, July 26--31, 2019, Proceedings, Part IV 21}.\hskip 1em plus 0.5em minus 0.4em\relax Springer, 2019, pp. 226--237.

\bibitem{soremekun2022software}
E.~Soremekun, M.~Papadakis, M.~Cordy, and Y.~L. Traon, ``Software fairness: An analysis and survey,'' \emph{arXiv preprint arXiv:2205.08809}, 2022.

\bibitem{chen2024fairness}
Z.~Chen, J.~M. Zhang, M.~Hort, M.~Harman, and F.~Sarro, ``Fairness testing: A comprehensive survey and analysis of trends,'' \emph{ACM Transactions on Software Engineering and Methodology}, vol.~33, no.~5, pp. 1--59, 2024.

\bibitem{starke2022fairness}
C.~Starke, J.~Baleis, B.~Keller, and F.~Marcinkowski, ``Fairness perceptions of algorithmic decision-making: A systematic review of the empirical literature,'' \emph{Big Data \& Society}, vol.~9, no.~2, p. 20539517221115189, 2022.

\end{thebibliography}

\end{document}